\documentstyle[12pt]{article} 
\newcommand{\be}{\begin{eqnarray}} 
\newcommand{\ee}{\end{eqnarray}}
\newcommand{\bq}{\begin{eqnarray}} 
\newcommand{\eq}{\end{eqnarray}} 
\font\ttww=cmr10 scaled \magstep1  
\def\CcC{{\hbox{\ttww C\kern-.45em{\vrule height.65em width0.08em
depth-.04em\hskip.45em }}}} 
\def\RrR{{\hbox{\ttww I\kern-.17em{R}}}}
\def\IiI{{\hbox{\ttww I\kern-.19em{I}}}}
\font\wwtt=cmr7 scaled \magstep1
\def\rRr{{\hbox{\wwtt I\kern-.17em{R}}}}  
\begin{document} \baselineskip=15.75 pt 
\pagestyle{empty} 
\vspace{.3 cm}
\begin{center} 
{\Large\bf Error Avoiding Quantum Codes}\\ 
\vspace{.5 cm}
{ PAOLO ZANARDI$^{1}$, and MARIO RASETTI$^{2}$}\\ 
\vspace{.3 cm}
{\sl $^{1}$ ISI Foundation, Villa Gualino, Viale Settimio Severo, Torino}\\
{\sl $^{2}$Dipartimento di Fisica and Unit\`a INFM, Politecnico di Torino,
}\\ 
{\sl Corso Duca degli Abruzzi 24, I-10129 Torino, Italy} \\ 
\end{center}
\pagestyle{plain} 
\vspace{.4 cm} 
\noindent{\small\bf Abstract}
{\small The existence is proved of a class of open quantum
systems that admits a linear subspace ${\cal C}$ of the space of states such
that the restriction  of the dynamical semigroup to the states built over $\cal
C$ is unitary. Such subspace allows for error-avoiding (noiseless) enconding of
quantum information.}\\ 
\medskip
\begin{flushleft} 
\end{flushleft} 
\vspace{.3 truecm} 
\begin{flushleft} {\bf 1.
Introduction} 
\end{flushleft} 
\vspace{.2 truecm} 
In this letter we deal with the general question, in the frame of the
mathematical theory of open quantum systems, whether a subset of the state
space of a given open system {\sl S} within the environment {\sl E} exists,
unaffected by the coupling of {\sl S} with {\sl E}.  Such a challenging
question raises with special enphasis in the area of {\sl quantum computation}
($QC$) \cite{QC}, where it finds strong motivations. 
{\sl QC} aims to construct computational schemes, based on
quantum features,  more efficient ({ e.g.} exponentially faster)
than classical algorithms \cite{BEVA}. Quantum computation  differs from
classical computation in that, whereas in the latter a Turing-Boole state is
specified at any time by a single integer, say $n$, written in binary form, the
generic state $|\psi \rangle$ of a quantum computer is a superposition of 
states $|n\rangle$ in some appropriate Hilbert space ${\cal H}$, each of which 
can be thought of as corresponding to a classical boolean state; 
$\displaystyle{ |\psi\rangle\, =\sum_{n=00\cdot\cdot\cdot 0}^{11\cdot\cdot
\cdot 1} c_{n}|n\rangle}.$ The features of $|\psi\rangle$ are described by the 
probability amplitudes
$c_{n}$. The higher potential efficiency one may expect of quantum with respect
to classical computation is ascribable just to characteristically quantum
mechanical properties, such as interference (the phases of the $c_{n}$'s play a
role), entanglement (some of the quantum states of a complete system do not
correspond to definite states of its constituting parts), von Neumann state
reduction (a quantum state cannot be observed without being irreversibly
disturbed), which are absent in classical computers. Moreover,
quantum information processing is inherently parallel, due to the linear
structure of state space and of dynamical evolution. For example, the quantum
Turing machine proposed by Deutsch \cite{DD1} consists of a unitary evolution
from a single initial state encoding {\sl input} data to a final state encoding
the {\sl output}. 
As in Turing's scheme, the initial state encodes information
on both the input and the $''$program$''$. 
It is therefore clear that in the physical implementation of $QC$ maintaining
quantum coherence (namely the phase relationship between the $c_n$'s)
in any computing system is an essential requirement in order
to take advantage of its specific quantum mechanical features. 
On the other hand any real system
unavoidably interacts with some environment, wich, typically, consists of a
huge amount of uncontrollable degrees of freedom. Such interaction causes a
corruption of the information stored in the system as well as errors in
computational steps, that may eventually lead to wrong outputs. One of the
possible approaches to overcome such difficulty, in  analogy with classical
computation, is to resort to redundancy in encoding information, by means of
the so-called (quantum) {\sl error correcting codes} (ECC). In these schemes 
\cite{ERROR}
information is encoded in linear subspaces $\cal C$ (codes) of the system
Hilbert space in such a way that $''$errors$''$ induced by the interaction 
with the enviroment can be detected and corrected. Of course, detection of
correctable errors has to be carried over with no gain of {\sl which-path}
information about the actual system state; otherwise this would result in a
further source of loss of coherence. The ECC approach appears then to aim 
to an {\sl active} stabilization of quantum states by conditionally carrying 
on suitable quantum operations \cite{KRA}.
 The typical system considered in the ECC literature is the
$N$-{\sl qubit register} $ R$ made of $N$ replicas of a two-level system 
$S$ (the qubit) where each qubit of $ R$ is assumed to be coupled with an
independent environment. We shall prove here that, by relaxing the latter
assumption, one can identify a class of open quantum systems which admit linear
subspaces ${\cal C}$ such that the restriction to ${\cal C}$ of the dynamics is
unitary. Quantum information encoded in such subspaces is therefore preserved,
thus  providing a strategy to maintain  quantum coherence. 
The approach to the decoherence problem suggested by our results \cite{ZARA}
is, in a sense, complementary to EC, in that it consists in a {\sl passive}
stabilization of quantum information.
For this reason,  subspaces ${\cal C}$ will be referred to
as {\sl Error Avoiding Codes} ($EAC$). 
\begin{flushleft}
{\bf 2. { Outline }}
\end{flushleft}
In this paper, without loss of generality  \cite{LIN}, 
we shall describe the quantum dynamics of a (open) 
system $S$ in terms of marginalization of the dynamics associated  
to a one-parameter unitary group $\{U_t\}_{t\in\rRr}$ of transformations
acting on an enlarged Hilbert space (system plus {\sl environment}).
Even though  this description is by no means unique, we assume that the form
of the generator (Hamiltonian) of the dynamical group  is dictated
by physical considerations \cite{ZARA}.
 The component of the Hamiltonian that
induces a non-trivial mixing  of the system and environment degrees of freedom
will, as usual, be referred to as the interaction Hamiltonian $H_I$.
In sect. 3 after  defining an EAC  $\cal C$ 
as a subspace with unitary marginal dynamics
we characterize it (Lemma 3.1)  by the simple property that   
$H_I$  restricted to $\cal C$ should be the identity on the system space.
The  simplest -- but physically important --
 example is provided by the simultaneous eigenspaces (if any)
of the whole set of system operators appearing in $H_I$
(Theorem 3.1). 
Such a condition can be implemented in a less trivial way by means
of the reducible structure of the system Hilbert space considered
as a representation space  of a  group $\cal G$
or of a Lie algebra   ${\cal A}_S$ (Theorems 3.2, 3.3).
In the former case $\cal G$ is required to be a symmetry group for
$H_I,$ in the latter the allowed interaction operators ({\sl error
generators}) have to belong to ${\cal U}({\cal A}_S)$.
By imposing that both    module-structures are present and compatible
(the representatives of  elements of ${\cal A}_S$ are $\cal G$-invariant)
one can identify (Theorem 3.4) a whole class of EAC's as the {\sl singlet sector}; 
direct sum of the 
one-dimensional submodules of a semisimple (dynamical) 
Lie algebra ${\cal A}_S$. 
   In sect. 4  we consider the case of a {\sl quantum register}
   $ R$ defined as the collection of $N$ replicas of a $d$-dimensional
quantum system ({\sl cell}) $C.$
Assuming that the error generators $\{S^{(i)}_\lambda\}\,(i=1,\ldots,N)$
 of each 
cell are coupled in a
 replica-symmetric way to a common environment, one finds that the 
marginal dynamics of $ R$ is described in terms of the $N$-fold
tensor representation $\phi_N$ of the {\sl dynamical algebra} ${\cal A}_S$
[isomorphic to $\mbox{sl}(d,\,\CcC)]$ 
spanned
by the $S^{(i)}_\lambda$'s. Theorem 3.4 holds because
$\phi_N$ is compatible with the natural action of the symmetric group
${\cal S}_N$. 
\begin{flushleft}
{\bf 3. Error Avoiding Codes }
\end{flushleft}
\vspace{.2 truecm}
Let ${\cal H}_\alpha,\, \mbox{dim}{\cal H}_\alpha=d_\alpha,\,(\alpha=E,S))$ be 
finite dimensional Hilbert spaces. The quantum system associated to ${\cal H}_S$
(${\cal H}_E$) will be referred to as the {\sl system} (respectively, the {\sl 
environment}). The set of non-negative hermitian operators on Hilbert space 
$\cal H$ with trace one will be denoted by ${\cal S}({\cal H})$; its elements
will be referred to as {\sl states}. 
 ${\cal S}({\cal H})$ is the convex hull of the set of {\sl pure states}
\begin{eqnarray}
{\cal S}_P({\cal H}_S)\doteq\{\rho\in{\cal S}({\cal H}_S)\,\colon\,
\rho^2=\rho\}\cong {{\cal H}_S}/{{\rm U}(1)} \; .
\label{pure}
\end{eqnarray}
We assume the quantum system associated 
with ${\cal H}_{SE}={\cal H}_S\otimes{\cal H}_E$ to be {\sl closed}, {\sl i.e.} 
its dynamics to be generated by a hermitian operator $H_{SE}\in\mbox{End}
({\cal H}_{SE})$. The time evolution of any state $\rho\in{\cal S}({\cal 
H}_{SE})$ is given by $\rho\rightarrow\rho_t\doteq U_{SE}(t)\,\rho\,
U^\dagger_{SE}(t)$, where $U_{ES}(t)\doteq \exp (-i\,t\,H_{SE}),\,(t\in\RrR)$ is the 
one-parameter unitary group generated by $H_{SE}$. 
The {\sl marginal} dynamics 
on ${\cal H}_S$ (conditional to the initial preparation $\rho_E\in{\cal S}
({\cal H}_E)$) is given by 
\begin{eqnarray}
{\cal E}_t^{\rho_E}\colon \; {\cal S}({\cal H}_S)\rightarrow{\cal S}({\cal 
H}_E)\; \colon\; \rho\rightarrow \mbox{tr}^E \left ( U_{SE}(t)\,\rho\otimes
\rho_E\,U_{SE}^\dagger(t)\right )\; .
\end{eqnarray}
The {\sl dynamical semigrup} $\{{\cal E}_t\}_{t\ge 0}$ does not leave invariant 
the set of { pure states}.
This  a characteristic  quantum phenomenon known as {\sl decoherence}. It reflects the fact
that the system-environment interaction {\sl entangles} the degrees of freedom
of $S$ with those of $E$ in such a way that, despite unitarity (which does
indeed preserve purity of the overall joint state) each of the two subsystems
has no longer a (pure) state of its own: the two subsystem have became {\sl
inseparable} \cite{PER}. From the point of view of quantum information this
amounts to a corruption of the initial state.\\ 
For $\cal C$ a $d_{{\cal C}}$-dimensional linear subspace of ${\cal H}_S$, 
we denotes by ${\cal A}({\cal 
C})$ the subalgebra of $\mbox{End}({\cal H}_S)$ leaving $\cal C$ invariant:  
${\cal A}({\cal C})\doteq \{X\in\mbox{End}({\cal H}_S)\,\colon\,
X\,{\cal C}\subset\cal C\}$. \\ 
{ {DEFINITION 3.1.} \label{def1} } {\it A linear subspace ${\cal C}\neq \{0\}$ of $ 
{\cal H}_S,$ is  an {\it error avoiding code}  ($EAC$) iff \\ 
i)  $\exists\,H_S\in {\cal A}({\cal C})$, $H_S,$ hermitian, is such that, $\forall
\rho_E\in{\cal S}({\cal H}_E)$, $\rho\in{\cal S}({\cal C})\Rightarrow {\cal
E}_t^{\rho_E}(\rho) =e^{-i\,t\,H_S}\,\rho\,e^{i\,t\,H_S} (\forall t\in\RrR)$.\\ 
ii) $\cal C$ is maximal ({\it i.e.} it is not a proper subspace of any space 
for which i) holds).} \\ 
Each state in $\cal C$ will be referred to as {noiseless}. \\ 
Definition 3.1 of $EAC$ can be summarized in terms of commutativity ($\forall t\in\RrR$)
of the following diagram: 
\def\normalbaselines{\baselineskip20pt\lineskip3pt\lineskiplimit3pt} 
\def\mapright#1{\smash{\mathop{\longrightarrow}\limits^{#1}}}
\def\mapup#1{\Big\uparrow\rlap{$\vcenter{\hbox{$\scriptstyle#1$}}$}} 
\begin{eqnarray}
\matrix{{\cal S}({\cal H}) & \mapright{{\cal E}_t} & {\cal S}({\cal H}) \cr 
        \mapup{\iota} && \mapup{\iota} \cr 
        {\cal S}_P ({\cal C}) & \mapright{{\rm Ad}(U_t)} & 
        {\cal S}_P ({\cal C}) \cr 
        \mapup{\cong} && \mapup{\cong} \cr 
        {\cal C}/{\rm U}(1) & \mapright{U_t} & {\cal C}/{\rm U}(1) \cr} 
        \nonumber    
\end{eqnarray} 
Here $\cong$ is the isomorphism defined in equation (\ref{pure}) and $\iota$ is the canonical 
inclusion map.\\
{\it{Remark 1.}}
The eigenstates of $H_S$ in $\cal C$ are stationary states
(i.e. ${\cal E}_t(\rho)=\rho\, ,\forall t\in\RrR$).\\
${\cal C}\neq\{0\}$ means that there exists a set of initial
preparations for which no information loss occurs. Since the minimal system
which permits useful enconding of quantum information is a two-level system
({\sl qubit}), an $EAC$ has use in $QC$ if $\mbox{dim}\,{\cal C}>1$. \\ 
The Hamiltonian $H_{SE}$ has the form
\begin{eqnarray}
H_{SE}= H_S\otimes\IiI_E+\IiI_S\otimes H_E + H_I \; ,
\end{eqnarray}
where $H_S$ ($H_E$) is an hermitian operator on ${\cal H}_S$ (respectively,
${\cal H}_E$) and $H_I$, hermitian, acts, for an arbitrary state, in a non
trivial way on both factors of the tensor product space ${\cal H}_{SE}$. The
following lemma states a sufficient and necessary condition for $EAC$'s:\\
{ {LEMMA 3.1.}} {\sl A linear subspace ${\cal C}\subset {\cal H}_S$ is an  
$EAC$ iff \\ 
i) $H_S\in{\cal A}( {\cal C})$, \\ ii) $H_I|_{{\cal C}} = \IiI_{\cal C}\otimes E({\cal 
C}),\, (E({\cal C})=E^\dagger({\cal C})\in\mbox{End}({\cal H}_E))$. }\\
{\it Proof}\\
We first show that i) and ii) are sufficient conditions. Let $\{|\phi_k\rangle\}$ be
an orthonormal set of eigenvectors of the hermitian operator $\tilde H_E \doteq
H_E+ E({\cal C})$, and $\{\tilde \epsilon_k\}$ the corresponding set of
eigenvalues. Any $\rho_E\in{\cal S}({\cal H}_E)$ can written in the form
$\rho_E=\sum_{k,h} R_{kh} |\phi_k\rangle\langle\phi_h|$, where $R$
 is a hermitian non-negative matrix of rank $d_E$ and trace one
with complex matrix elements $R_{kh}$. For $\rho \in{\cal S}({\cal C})$,
\begin{eqnarray}
{\cal E}_t^{\rho_E}(\rho) &=& \mbox{tr}^E\left ( U_{SE}(t)\,\rho\otimes\rho_E
\,U^\dagger_{SE}(t)\right )\nonumber \\
&=&\sum_{kh} R_{kh}\,  \mbox{tr}^E \left ( e^{-i\,t\,H_S}\,\rho\,e^{i\,t\, 
H_S}\otimes e^{-i\,t\,(\tilde \epsilon_k-\tilde \epsilon_h)}
|\phi_k\rangle\langle\phi_h|\right) \nonumber \\ 
&=& e^{-i\,t\,H_S}\,\rho\,e^{i\,t\, H_S} \sum_k R_{kk} =e^{-i\,t\,H_S}\,\rho
\,e^{i\,t\,H_S} \; , 
\end{eqnarray}
in that $\mbox{tr}^E \left (|\phi_k\rangle\langle\phi_h|\right )=\delta_{hk}$, 
and $\sum_k R_{kk}=1$.\\
Suppose now that $\cal C$ is an $EAC$.  Expanding the identity at point i) of 
Definition 3.1 up to the first order in $t$ one finds $[\tilde H(\rho_E),\,\rho 
]=0, \,\forall\rho\in{\cal S}({\cal C})$, where $\tilde H(\rho_E)\doteq  
\mbox{tr}^E\left (  \rho_E\, H_{SE}^\prime  \right )$, and $H_{SE}^\prime\doteq 
H_{SE}-H_S\otimes\IiI_E$. From the (manifest) commutativity of $\tilde H_{SE}$ 
with all the states of $\cal C$ ensues that $\tilde H_{SE}|_{\cal C}= 
\lambda(\rho_E)\, \IiI_{\cal C}$. Moreover, since this property holds for 
all $\rho_E\in{\cal S}({\cal 
H}_E)$, one has $\langle \phi_i|\, H^\prime_{SE}\,|\phi_i\rangle =\lambda_i\
\IiI_{\cal C},\, \forall |\phi_i\rangle\in{\cal H}_E$. It follows from this 
latter relation that $\langle \phi_i|\,H_{SE}^\prime\,|\phi_{i^\prime}\rangle=
\lambda_{i i^\prime}\,\IiI_{\cal C}$. Therefore the spectral resolution of 
$H_{SE}^\prime$ finally reads
\begin{eqnarray}
H_{SE}^\prime &=& \sum_{j j^\prime,i i^\prime} |\psi_j\rangle\otimes|\phi_i
\rangle \langle\psi_j|\langle\phi_i|\,H_{SE}^\prime\,|\phi_{i^\prime}\rangle|
\psi_{j^\prime}\rangle \langle\psi_{j^\prime}|\otimes\langle \phi_{i^\prime}| 
\nonumber \\ 
&=& \sum_j|\psi_j\rangle\langle \psi_j|\otimes \sum_{i i^\prime}\lambda_{i 
i^\prime} |\phi_i\rangle\langle\phi_{i^\prime}|= \IiI_{\cal C}\otimes E \; ,
\label{H'SE} 
\end{eqnarray}
for some $E\in\mbox{End}({\cal H}_E)$. Here $\{|\psi_j\rangle\}_{j=1}^{d_{\cal 
C}}$ ($\{|\phi_i\rangle\}_{i=1}^{d_E}$) is a orthonormal basis of ${\cal C}$ 
(respectively, ${\cal H}_E$). The {\sl r.h.s.} of eq. (\ref{H'SE}) shows that 
$H_{SE}-H_S\otimes\IiI_E,$ restricted to $\cal C$ acts trivially on the system 
Hilbert space, as was to be proven. \hfill$\Box$\\
{\it{Remark 1.}} Suppose a unitary $U\in\mbox{End}({\cal H}_S)$ exists  
such that $\mbox{Ad}\,U(H_{SE})\doteq U\,H_{SE}\,U^\dagger$ satisfies the 
hypothesis of Lemma 3.1 with respect to subspace $\cal C$. Then $U^\dagger\,
{\cal C}$ is an $EAC$.\\ 
The physical meaning of Lemma 3.1 is quite transparent: the states over $\cal C$ 
do not suffer any decoherence in that they are all affected by the environment in the same 
way.\\
The general form of the interaction Hamiltonian $H_I$ is
\begin{eqnarray}
H_I=\sum_{\lambda\in\Lambda} S_\lambda\otimes E_\lambda \; ,
\label{Hint}
\end{eqnarray}
where 
$X_\lambda\in\mbox{End}({\cal H}_X), X=S,E,$ and $\Lambda$ is a suitable 
(finite) index set. 
The operators $\{S_\lambda\}$ will be referred to as {\sl error generators}.
Lemma 3.1 basically asserts that $\cal C$ is an $EAC$ 
iff $H_S\in{\cal A}({\cal C})$ and the $S_\lambda$'s belong to the subalgebra 
${\cal A}_1({\cal C})\subset {\cal A}({\cal C})$ of operators with restriction 
to ${\cal C}$ proportional to the identity. Notice that ${\cal A}_1({\cal C})$ 
contains the ideal ${\cal A}_0({\cal C})$ of those operators in ${\cal A}_1
({\cal C})$ which annihilate $\cal C$. If the error generators belong to 
${\cal A}_0({\cal C})$ the dynamics on ${\cal C}\otimes{\cal H}_E$ coincides 
with that generated by the free Hamiltonian.\\
The simplest case in which Lemma 3.1 provides an $EAC$ is described in the 
following\\
{{THEOREM 3.1.}}
{\it Let $\{ S_\lambda\}_{\lambda\in\Lambda}$ and $H_S$ form a commutative family of 
hermitian operators. If ${\cal C}$ is a maximal common eigenspace of the 
$S_\lambda$'s, then $\cal C$ is an $EAC$.\\}
{\it Proof}\\
Let $\sigma_\lambda,\, (\lambda\in\Lambda)$ be the set of 
$S_\lambda$-eigenvalues, then one has $H_I|_{{\cal C}}=\sum_{\lambda\in
\Lambda} \sigma_\lambda \IiI_{\cal C}\otimes\,E_\lambda= \IiI_{\cal C}\otimes \sum_{\lambda\in 
\Lambda} \sigma_\lambda\, E_\lambda\doteq\IiI_{\cal C}\otimes E({\cal C})$. Since 
$\cal C$ is maximal and $H_S$ commutes with the $S_\lambda$'s, then $H_S\in{\cal A}({\cal C}),$
 and the thesis follows  from Lemma 3.1. \hfill$\Box$\\
Let now $\cal G$ be a group, $\Phi$ a unitary representation of $\cal G$ on ${\cal 
H}_S$. ${\cal H}_S$, considered as a ${\cal G}$-module, has the decomposition, 
in terms of irreducible ${\cal G}$-submodules
\begin{eqnarray}
{\cal H}_S=\bigoplus_{j\in{\cal J}} n_j\, {\cal H}_j \; ,
\label{Gsplit}
\end{eqnarray}
where ${\cal J}$ is a label set for the $\cal G$-irreps, $\{{\cal H}_j\}_{j
\in{\cal J}}$ is the set of irreducible submodules of $\cal G$, and the integers 
$\{n_j\}_{j\in{\cal J}}$ are the corresponding multiplicities.
Suppose $\exists j_0\in{\cal J}$ such that $n_{j_0}(\Phi)=1$, and let $\cal C$ 
be the corresponding submodule; then\\
{{THEOREM 3.2.}}
{\it If the $\{S_\lambda\}$'s in equation (\ref{Hint}) are }$\mbox{Ad}\, 
\Phi({\cal G})${\it -invariant and  $H_S\in{\cal A}(\cal C),$ then $\cal C$  is 
an $EAC$. }\\
{\it Proof}\\
Since the $S_\lambda$'s transform according to the identity representation of 
$\cal G$, they can couple only submodules corresponding to equivalent 
representations. Therefore it follows from $n_{j_0}(\Phi)=1$ that $ S_\lambda
\in{\cal A}({\cal C})\, ( \lambda\in\Lambda)$. Hence the $S_\lambda$'s 
commute with all operators of the $\cal G$-irrep labelled by $j_0$, and 
one obtains  -- from Schur's lemma -- that $S_\lambda|_{{\cal C}} \sim \IiI_{\cal C}\, 
(\lambda\in\Lambda)$. The thesis follows from Lemma 3.1. $\hfill\Box$\\
Let us suppose now that the error generators  belong to some  
representation $\phi\colon {\cal A}_S \rightarrow \mbox{gl}({\cal H}_S)$
of a Lie algebra ${\cal A}_S$ ({\sl dynamical algebra}). $\phi$ turns 
${\cal H}_S$ into an ${\cal A}_S$-module that has a decomposition analogous to 
equation (\ref{Gsplit}) ($\cal J$ being now a label set for the ${\cal 
A}_S$-irreps). \\
{{THEOREM 3.3.}}
{\it Let ${\cal C}$ be the direct sum over a maximal set of equivalent  
one-dimensional ${\cal A}_S$-submodules. Suppose ${\cal C}\neq\{0\}$ and 
$H_S\in{\cal A}({\cal C})$; then ${\cal C}$ is an $EAC$.}\\
{\it Proof}\\
Since $\cal C$ is spanned by ${\cal A}_S$-singlets and $\{S_\lambda\}\subset 
\phi({\cal A}_S)$ one has $ S_\lambda\,|\psi\rangle =\sigma_\lambda\,|\psi 
\rangle,$ where the  $\sigma_\lambda$ are c-numbers. Therefore the assumption 
of Lemma 3.1 holds with $E({\cal C})=\sum_{\lambda\in\Lambda} \sigma_\lambda 
\,E_\lambda$. \hfill$\Box$ \\
{\it {Remark 1.}}
When ${\cal A}_S$ is semisimple, then the $\sigma_\lambda$'s are necessarily 
zero, and all the one-dimensional irreps are equivalent.\\
{\it {Remark 2.}}
When ${\cal A}_S$ is abelian all the irreps are one-dimensional. The subspaces 
corresponding to the direct sum over a maximal set of equivalent irreps are 
{\sl weight} spaces.\\
{\it {Remark 3.}}
Theorem 3.3 still holds if the error generators  belong to $\phi({\cal 
U}({\cal A}_S))$, where ${\cal U}({\cal A}_S)$ denotes the {\sl universal 
enveloping algebra} of ${\cal A}_S$. \\
The Lie-algebra representation  $\phi$ is {\sl compatible} ({\sl i.e.} $\mbox
{Ad}\,\Phi({\cal G})$-invariant) with the action of the group $\cal G$ iff 
$\Phi(g)\,X\,\Phi^\dagger(g)=X,\, \forall g\in{\cal G},X\in\phi({\cal A}_S)$. 
In this case, when ${\cal A}_S$ is semisimple,
 the multiplicities of the ${\cal A}_S$-irreps ($\cal G$-irreps) 
appearing in the decomposition of $\phi$ ($\Phi$) are but the dimensions of
the $\cal G$-irreps (${\cal A}_S$-irreps) entering the decomposition of $\Phi$ 
($\phi$). In particular this means that the subspace $\cal C$ obtained as 
direct sum over the one-dimensional ${\cal A}_S$-submodules of $\phi$ ({\sl singlet sector})
appearing in the decompostion of $\phi$ is a $\cal G$-module which enters with 
multiplicity one in the decomposition of $\Phi$.  \\
{{THEOREM 3.4.}}
{\it Let $\cal C$ be the singlet sector of the $\cal G$-compatible Lie-algebra 
representation $\phi$ of ${\cal A}_S.$ If \\ 
i) the error generators are} $\mbox{Ad}\,\Phi({\cal 
G})${\it -invariant, \\ 
ii) $H_S\in {\cal A}({\cal C})$, \\ 
then $\cal C$ is an $EAC$.}\\
{\it Proof}\\
The singlet sector corresponds to a $\cal G$-irrep appearing in the $\Phi$ 
decomposition with multiplicity one. The thesis follows 
from Theorem  3.2.\hfill$\Box$\\
{\it{Remark 1.}}
$\phi({\cal U}({\cal A}_S))$ is $\mbox{Ad}\,\Phi({\cal G})$-invariant in that 
it is generated by $\IiI$ and $\phi({\cal A}_S)$.  \\
{\it{Remark 2.}}
If $[H_S,\,\phi({\cal A}_S)]=0$ or $H_S\in\phi({\cal U}({\cal A}_S))$,
the condition $H_S\in{\cal A}({\cal C})$ is fulfilled.
\begin{flushleft}
{\bf 4. {Quantum Registers}}
\end{flushleft}
In this section the physically relevant notion of {\sl register} 
is introduced, in analogy with the case of classical computation. \\ 
{{DEFINITION 4.1.}}
{\it A $d$-dimensional (quantum) cell $C$ is a quantum system associated to 
a Hilbert space ${\cal H}_C\cong \CcC^d$. A (quantum) register with $N$ cells 
is a quantum system given by $N$ replicas of $C$.  $R$ is associated 
with ${\cal H}_R={{\cal H}_C}^{\otimes\,N}$. }\\
The register self-hamiltonian will be denoted as $H_R$. The register Hilbert 
space ${\cal H}_R$ is a natural ${\cal S}_N$-module.
Let $\{|\psi_j\rangle\}_{j=1}^d$ be a basis of ${\cal H}_C$; one can define
$\sigma\cdot \otimes_{k=1}^N|\psi_{j_k}\rangle= \otimes_{k=1}^N|
\psi_{j_{\sigma(k)}}\rangle,\,(\forall \sigma\in{\cal S}_N)$. 
The latter formula defines, by linear extension, a representation $\Phi_N$ of
${\cal S}_N$ on ${\cal H}_R$. The operators compatible with this ${\cal 
S}_N$-action lie in the symmetric subspace of $\mbox{End}\,({\cal H}_R)\cong
\mbox{End}^{\otimes\,N}\,{\cal H }_C$. If each cell of $R$ is coupled with the 
(common) environment $E$ by a ${\cal S}_N$-invariant interaction, one has
\begin{eqnarray}
H_I=\sum_{i=1}^N\sum_{\lambda\in\Lambda} S_\lambda^{(i)}\otimes E_\lambda
\in\mbox{End}\,({\cal H}_R\otimes{\cal H}_E) \; ,
\label{Hreg}
\end{eqnarray}
where $S^{(i)}_\lambda\in\mbox{End}\,({\cal H}_R),\,(i=1,\ldots,N,\,\lambda\in 
\Lambda)$ acts as $S_\lambda$ in the $i$-th factor of the tensor product 
${\cal H}_R$, and as the identity in the other factors.\\
Notice that the register-environment interaction (\ref{Hreg}) involves only
the {\sl coproduct} operators $\Delta^{(N)}(S_\lambda)\doteq\sum_{i=1}^N S_ 
\lambda^{(i)}$. If $\phi\colon{\cal A}_S\rightarrow \mbox{gl}({\cal H}_C)$ is 
a representation of the Lie algebra ${\cal A}_S$ in ${\cal H}_C$, then 
$\Delta^{(N)}\circ \phi\colon {\cal A}_S\rightarrow\mbox{gl}({\cal H}_R)$ is 
the $N$-fold tensor product of $\phi$ and will be denoted as $\phi_{N}$. 
An important role in physical applications is played by the case in which 
${\cal A}_S\cong\mbox{sl}(d,\CcC)$ and $\phi$ is the defining representation.\\
{{ THEOREM 4.1}}
{\it Let the quantum register $R$ be coupled with the environment $E$
by the Hamiltonian given by equation (\ref{Hreg}), where the interaction 
operators $S_\lambda$ belong to the defining representation $\tilde\phi$ of} 
$\mbox{sl}(d,\CcC)$ {\it in ${\cal H}_C$.  Let ${\cal C}_N$ 
be the singlet sector 
of $\tilde \phi_N$. If $H_R\in{\cal A}( {\cal C}_N)$ then ${\cal C}_N$ 
is an $EAC$.}\\
{\it Proof}\\
>From Theorem 3.4, letting ${\cal A}_S=\mbox{sl}(d,\,\CcC)$, ${\cal G}={\cal 
S}_N$, $\phi=\tilde \phi_N$, and $\Phi=\Phi_N.$ \hfill$\Box$\\
{\it{Remark 1.}} Remarks 1. and 2. of Theorem 3.4 imply immediately that 
in the latter proposition the error generators are allowed to belong to 
$\tilde \phi_N({\cal U}(\mbox{sl}(d,\,\CcC)))$ as well. In this case the latter 
subspace coincides with the whole space of ${\cal S}_N$-invariant operators. \\
\medskip
\begin{flushleft}
{\bf 5. Conclusions}
\end{flushleft}
In this paper we introduced the notion of Error Avoiding Quantum Code
as the subspace $\cal C$ of the Hilbert space of an open quantum system $S$
enbedded in an environment $E,$
in which quantum coherence is preserved.
Formally this means that the dynamical (one-parameter) semigroup of $S$ 
restricted 
to initial data in ${\cal S}({\cal C})$ is given by a (one-parameter) group of
unitary transformations. 
We proved a number of theorems which relate the existence of an $EAC$ to the
(dynamical)  algebraic structure of the interaction Hamiltonian coupling $S$ 
and $E.$ In particular we discussed the case of a quantum register 
symmetrically coupled with the environment. 
>From the broader point of view of the theory of open quantum systems, our
results provide a systematic way of building non-trivial models in which, 
under quite generic assumptions, the  
unitary evolution of a subspace is allowed, even while the remaining part of the
Hilbert space gets strongly entangled with the environment. 
\medskip
\begin{flushleft}
{\bf Acknowledgements}
\end{flushleft}
One of the authors (P.Z.) thanks C. Calandra
for providing  hospitality at the University of  Modena,
and Elsag-Bailey for financial support.
 
\end{document}